\documentclass[aps,prl,twocolumn,groupaddress,amssymb,amsmath,nofootinbib]{revtex4}

\usepackage{color}
\usepackage{graphicx}
\usepackage{epstopdf}
\usepackage{hyperref} 

\begin{document}

\preprint{}

\title{%Simple extensions of the standard model and 
Insensitive  Unification of Gauge Couplings}

\author{Radovan Derm\' \i\v sek}

\affiliation{Physics Department, Indiana University, Bloomington, IN 47405, USA}

%\email[]{dermisek@ias.edu}

        %\thanks{}
%\altaffiliation {}

%\date{\today}
\date{April 29, 2012}

\begin{abstract}

The standard model extended by three vector-like families with masses of order  1 TeV -- 100 TeV allows for unification of gauge couplings.
 The values of gauge couplings
 %,  top Yukawa coupling and  Higgs quartic coupling 
 at the electroweak scale are highly insensitive to fundamental parameters. The grand unification scale is large enough to avoid the problem with fast proton decay. The electroweak minimum of the Higgs potential is stable.  

\end{abstract}


\maketitle

%\section{ \label{sec:}}

%\subsection{}

%\subsubsection{}

%\subsection{Introduction}

%\subsubsection{Experimental results}

\noindent
{\bf Introduction.}
Our current understanding of elementary particles and their interactions is  described by the standard model  of particle physics (SM). It contains many free parameters, namely masses and mixing of quarks and leptons of the three families, the Higgs boson mass, and most importantly, the values of three gauge couplings  $\alpha_{1,2,3}$ that determine  the strengths of electromagnetic, weak, and strong interactions.
One of the main goals in particle physics is to understand values of these free parameters  from basic principles. 

Among the most elegant approaches is the idea of grand unification in which three gauge couplings $\alpha_{1,2,3}$, corresponding to three different symmetries of the SM, originate from a single gauge coupling associated with the symmetry of a grand unified theory (GUT)~\cite{PDG_GUTs}. 
%{\color{red} This idea is supported by the fact that quantum numbers of quarks and leptons in the SM are such that each family nicely fits into complete representations of the GUT symmetry, $\bf 10$ and $\bf \bar 5$ of $SU(5)$ or $\bf 16$ of $SO(10)$.} 
However, in the standard model the gauge couplings do not unify. Although through renormalization group (RG) evolution they run to comparable values at about $10^{14}$ GeV the mismatch is too large to be accounted for by GUT scale threshold corrections. In addition, the value of the unification scale is too low to satisfy limits on proton decay~\cite{PDG_GUTs}. Thus, in order to realize this idea, the SM has to be extended by additional particles.

In this letter we show that extending the standard model by three complete vector-like families (3VF) with masses of order  1 TeV - 100 TeV allows for unification of gauge couplings. Predictions for gauge couplings at the electro-weak (EW) scale are highly insensitive to fundamental parameters, and 
 ratios of  observed values  are to a large extent understood from the particle spectrum itself. The GUT scale can be sufficiently large to avoid the problem with fast proton decay, thus resurrecting simple non-supersymmetric GUT models.

The way this scenario works can be summarized in few steps. First,  extra 3VF make all gauge couplings asymptotically divergent which opens a possibility for 
a unification with large (but still perturbative) unified gauge coupling. Consequently, in RG evolution to lower energies gauge couplings run to the infrared fixed point.
Second,  the ratios of gauge couplings close to the infrared fixed point depend mostly on the particle content of the theory and they happen to be not far from the observed values. %at a chosen scale. 
Finally, the discrepancies between values of gauge couplings predicted from closeness to  infrared fixed point and corresponding observed values  can be fully explained by  threshold effects  from  masses of particles originating from 3VF. 
We note, that the first part is very similar to attempts to explain observed values of gauge couplings from infrared fixed point with 8 to 10 chiral families~\cite{Maiani:1977cg} (see also Refs.~\cite{Cabibbo:1982hy, Grunberg:1987sk}) before the number of chiral families and  values of gauge couplings were tightly constrained.

Allowing arbitrary new particles  provides many possibilities for gauge coupling unification. However, with arbitrary new particles the predictive power is typically lost and a  GUT embedding of such models is more complicated. On the other hand,
the addition of complete families, either  chiral (only left-handed or only right-handed particles with  given quantum numbers) or vector-like  (for each left-handed particle there is a right-handed particle with the same quantum numbers)
% to  the standard model
 represents some of the simplest  extensions of the SM (especially since we have no understanding of why there should be just three families in nature), that can be easily embedded into GUTs. 
 We will also argue that 
the predictive power of this scenario is comparable to  minimal supersymmetric  unification.
%, which predicts $\sim10\%$ too large $\alpha_3 (M_Z)$. This discrepancy is usually explained by $\sim3\%$  
%- 4\% 
%GUT scale threshold corrections, or by splitting  superpartner masses~\cite{PDG_GUTs}.

While the existence of extra chiral families  is unlikely given the current experimental situation, the vector-like families are poorly constrained~\cite{VFconstraints}.  This is mostly because a vector-like pair of fermions can have a mass which is not related to  their couplings to  the Higgs boson.  
Depending on the dominant decay mode, the limits on new vector-like fermions range from $\sim 100$ GeV to $\sim 500$ GeV.
Consequently, there is a vast literature on models  that contain extra vector-like fermions near the TeV scale. Examples include  attempts to explain the anomaly in   the forward-backward asymmetry of  the b-quark~\cite{Choudhury:2001hs, Dermisek:2011xu}, the muon g-2 anomaly~\cite{Kannike:2011ng}, and studies of effects of extra VFs in supersymmetric theories on gauge coupling unification~\cite{Babu:1996zv, BasteroGil:1999dx}, spectrum of superpartners~\cite{Kolda:1996ea},  or the Higgs mass~\cite{Martin:2009bg}, among many others.

\noindent
{\bf Adding three vector-like families.}
Consider the one-loop RG
%renormalization group 
equations (RGEs) for three gauge couplings
%, $\alpha_i = g_i^2 / 4\pi$
:
\begin{equation}
\frac{d \alpha_i}{dt} \; = \beta(\alpha_i) \; = \; \frac{\alpha_i^2}{2 \pi} \,  b_i,
\label{eq:RGE}
\end{equation}
where $t = \ln Q/Q_0$ with $Q$  representing the energy scale at which gauge couplings are evaluated, and the beta function coefficients in the SM are given by
\begin{equation}
 b_i \; = \; \left( \frac{1}{10}+\frac{4}{3}n_f , \;   - \frac{43}{6} +\frac{4}{3}n_f, \; -11 + \frac{4}{3}n_f\right),
 \label{eq:bi}
\end{equation}
where $n_f$ is the number of families.
For $n_f = 3$ we get the usual result, $b_i = (41/10, \; -19/6, \; -7)$, while with  extra 3VF, $n_f = 3 + 2\times 3 =9$
(a vector-like partner contributes in the same way),  we find $b_i = (121/10, \; 29/6, \; +1)$. Thus in the SM+3VF all three couplings are asymptotically divergent. 

\begin{figure}[t,h]
\includegraphics[width=2.6in]{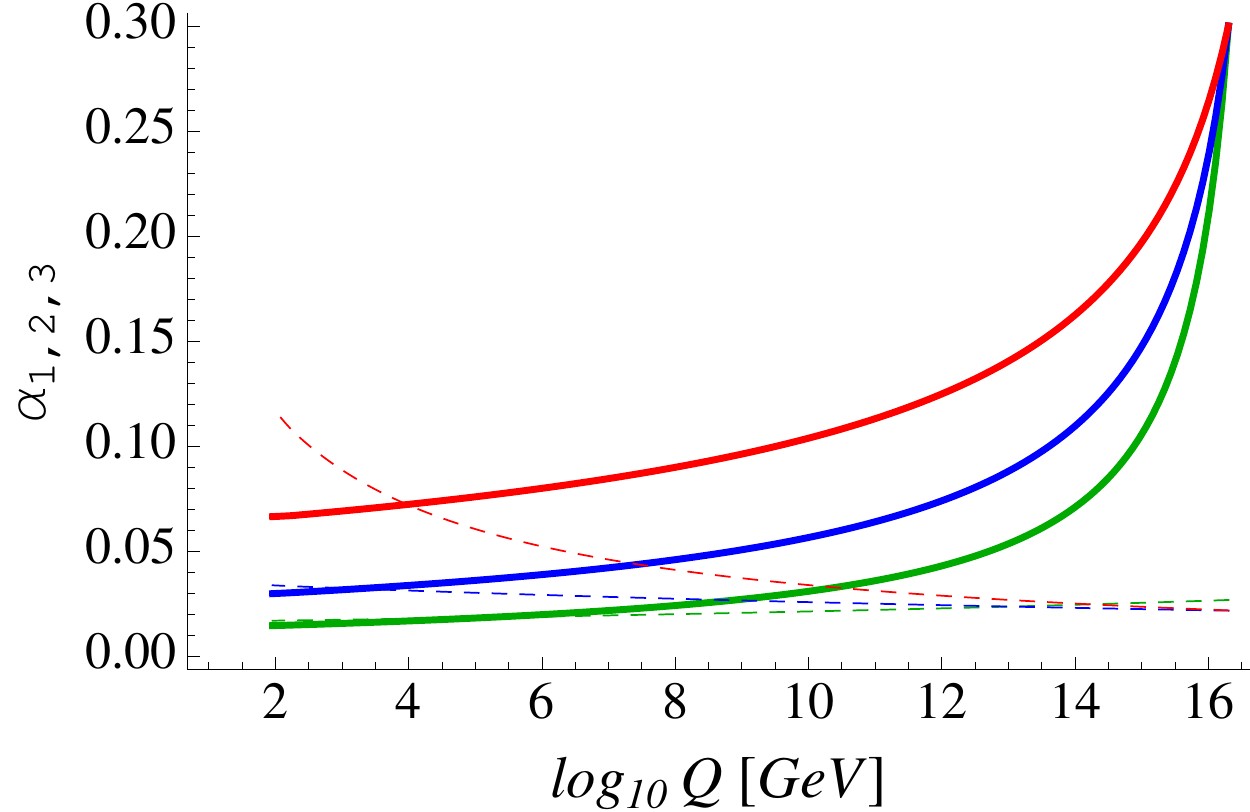}
\includegraphics[width=2.6in]{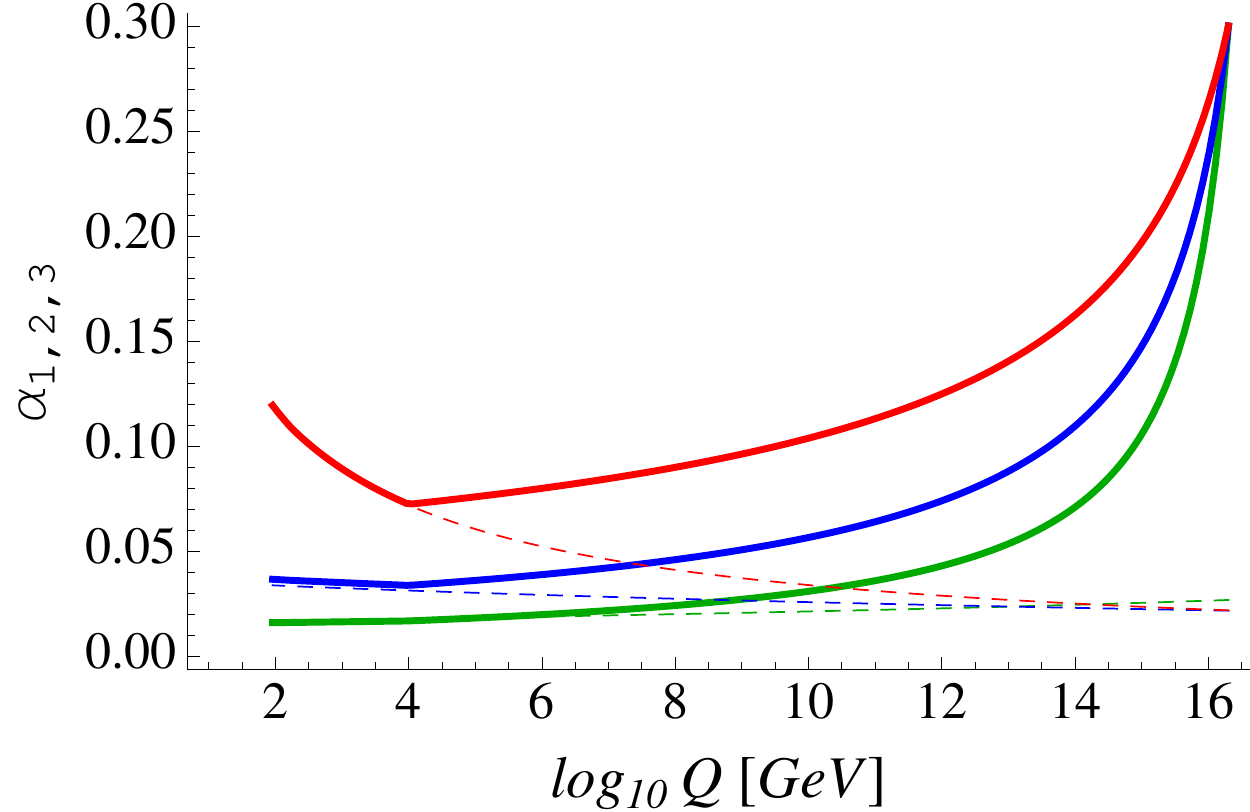}
\caption{RG running of gauge couplings: $\alpha_3$ (top solid line), $\alpha_2$ (middle solid line), and $\alpha_1$ (bottom solid line) in the SM extended by three vector-like families for $\alpha_G = 0.3$ at $M_G = 2 \times 10^{16}$ GeV.  Dashed lines in the same order  show  running of gauge couplings in the SM. Masses of  3VF are neglected in the top plot, and fixed to 10 TeV in the bottom plot. }
\label{fig:1}
\end{figure}

An example of the RG evolution of gauge couplings in the SM+3VF with unified gauge coupling $\alpha_G = 0.3$ at $M_G = 2 \times 10^{16}$ GeV is given in Fig.~\ref{fig:1}. The $M_G$ is chosen such that the limits on proton decay are satisfied.  Our choice also coincides with the scale of minimal supersymmetric unification and thus  allows for a direct comparison.
The evolution of gauge couplings in the SM, that will be useful for discussion of threshold corrections, is showed by dashed lines.  
As  input we use the  experimental central values of  $\alpha_{EM}^{-1} (M_Z) = 127.916$
% \pm 0.015$, 
and $\sin^2 \theta_W = 0.2313$,
% \pm 0.0002 $
which are related to $\alpha_{1,2} (M_Z)$ through 
% \begin{equation}
%\sin^2 \theta_W \equiv  \frac{\alpha'}{ \alpha_2 +  \alpha'} , \quad \quad \alpha_{EM} \; = \; \alpha_2 \, \sin^2 \theta_W,
%\end{equation}
$\sin^2 \theta_W \equiv  \alpha'/( \alpha_2 +  \alpha')$ and $ \alpha_{EM} \; = \; \alpha_2 \, \sin^2 \theta_W$,
where, using the $SU(5)$ normalization of hypercharge, $\alpha' \equiv  (3/5)  \alpha_1$; then 
$\alpha_{3} (M_Z) = 0.1184$; and the top quark mass $m_t = 173.2$ GeV. These together with  current experimental uncertainties can be found in Ref.~\cite{Nakamura:2010zzi}. 
We neglect Yukawa couplings of other fermions in the SM,  and for simplicity, we also assume that Yukawa couplings of 3VF are negligible. 
We further assume the Higgs boson mass $m_h = 125$ GeV which is in the middle of the currently allowed range~\cite{ATLAS:2012ae}. 
In all numerical results we use full two loop RGEs~\cite{Machacek:1983tz}. All particles with masses above $M_Z$ are integrated out at their mass scale. We include one-loop matching corrections for $m_t$ and $m_h$~\cite{Hambye:1996wb}.  
%(theta function approximation). 
However, in order to understand results,  approximate analytic formulas will be sufficient.

The one-loop RGEs, which are very good approximations for $\alpha_{1,2}$,  can be easily solved, and   gauge couplings at the EW scale can be written in terms of the GUT scale, $M_G$, and the unified gauge coupling, $\alpha_i (M_G) = \alpha_G$:
 \begin{equation}
 \alpha_i^{-1} (M_Z) \; = \;  \frac{b_i}{2 \pi}  \ln \frac{M_G}{M_Z} \; + \;  \alpha_G^{-1}  \; - \; T_i ,
 \label{eq:sol_1loop}
  \end{equation}
where $T_i$  are the threshold corrections that depend on masses of the extra vector-like fermions. They can be approximated
by the leading logarithmic corrections:
   \begin{eqnarray}
T_i  &\simeq &  \frac{1}{2 \pi}  \sum_{f}  b_i^f  \ln \frac{M_f}{M_Z} ,
\label{eq:Ti}
  \end{eqnarray}
where $b_i^f$ is the contribution  of a given fermion $f$, with mass $M_f$, to the corresponding beta function coefficient~\cite{Machacek:1983tz}.
Neglecting these  corrections  for a moment, we find that for large enough unification scale and sizable but still perturbative unified gauge coupling the first term in Eq.~(\ref{eq:sol_1loop}) dominates, $\alpha_i^{-1} (M_Z)  \simeq   (b_i/2 \pi)  \ln (M_G/M_Z)$ ,
%\begin{equation}
 %\alpha_i^{-1} (M_Z) \; \simeq \;  \frac{b_i}{2 \pi}  \ln \frac{M_G}{M_Z} ,
  %\end{equation}
  and the ratios of gauge couplings are completely fixed by ratios of beta function coefficients, $\alpha_i (M_Z)/  \alpha_j (M_Z)  \simeq b_j/b_i$.
 %  \begin{equation}
 %\frac{\alpha_i (M_Z)}{  \alpha_j (M_Z)} \; \simeq \;  \frac{b_j} {b_i} .
%\end{equation}
 Alternatively, in this 0-th order approximation, we obtain a {\it parameter-free prediction} for $\sin^2 \theta_W$:   
 \begin{equation}
\sin^2 \theta_W \equiv  \frac{\alpha'}{ \alpha_2 +  \alpha'} = \frac{b_2}{b_2+b'}=  0.193,
\end{equation}
where %using $SU(5)$ normalization of hypercharge, $\alpha' = (3/5)  \alpha_1$ and 
$b' \equiv  (5/3) b_1$, and  the numerical value corresponds to the SM+3VF. It is identical to the one obtained assuming 9 chiral families~\cite{Maiani:1977cg, Cabibbo:1982hy, Grunberg:1987sk}.
Although it is not a perfect match to the measured value, we will see that the discrepancy can be easily accommodated by taking into account  threshold corrections, $T_i$, and finite value of $\alpha_G$.
  
For $\alpha_3$, the one-loop RGE, Eq.~(\ref{eq:RGE}), is not a good approximation because of the accidentally small $b_3$ factor. The two-loop contribution to the beta function, $\alpha_3^3 B_3 /8\pi^2$, where $B_3 = -102 + (76/3) n_f = 126$ for SM+3VF~\cite{Machacek:1983tz}, is larger than the one-loop contribution for $\alpha_3 \gtrsim 0.1$ (the coupling is  perturbative, the 3-loop contribution is only a small correction for $\alpha_3 \simeq 0.1$).
Neglecting the one-loop contribution and threshold corrections, the RGE for $\alpha_3$ can be easily solved:
 \begin{equation}
 \alpha_3^{-2} (M_Z) \; \simeq \;  \frac{B_3}{4 \pi^2}  \ln \frac{M_G}{M_Z}  \; + \;  \alpha_G^{-2}  .
  \label{eq:al3_2loop}
 \end{equation}
 Again, for our  $M_G$ and $\alpha_G$,  the log term  dominates, $\alpha_3^{-2} (M_Z) \simeq   (B_3/4 \pi^2)  \ln (M_G/M_Z)$ ,
% \begin{equation}
 %\alpha_3^{-2} (M_Z) \; \simeq \;  \frac{B_3}{4 \pi^2}  \ln \frac{M_G}{M_Z} ,
 %\end{equation}
 and we obtain the second {\it parameter-free prediction}:
\begin{equation}
 \frac{\alpha_3^{2} (M_Z)}{\alpha_{EM} (M_Z)} \; \simeq \;  2\pi \frac{b_2+b'}{B_3}.
 \end{equation}
Numerically, for  $ \alpha_{EM}  (M_Z) = 1/127.916$, it predicts $\alpha_3  (M_Z) = 0.099$.
The 1-loop contribution can be added as an expansion in $\epsilon = 4 \pi b_3/B_3$: $\alpha_3 (M_Z) \to \alpha_3 (M_Z)/K$, where $K = 1+ (1/3) (\epsilon/\alpha_3 (M_Z)) - (1/12)(\epsilon/\alpha_3 (M_Z))^2 + \dots$. 
%\begin{equation}
%\alpha_3 (M_Z) \to \frac{\alpha_3 (M_Z)}{1+ \frac{1}{3} \frac{\epsilon}{\alpha_3 (M_Z)} - \frac{1}{12}\left(  \frac{\epsilon}{\alpha_3 (M_Z)}\right)^2 + \dots},
%\end{equation} 
It represents $\sim26 \%$ correction and leads to $ \alpha_3 (M_Z)  \simeq 0.073$ which is close to  0.066  obtained  using full two-loop RGEs, see Fig.~{\ref{fig:1} (top). 
The discrepancy from the  observed value can be accommodated by the threshold correction, $T_3$, which is well approximated by Eq.~(\ref{eq:Ti}) and can be added to $ \alpha_3^{-1} (M_Z)$ as in Eq.~(\ref{eq:sol_1loop}).

\noindent
{\bf Threshold corrections.} Assuming a common mass, $M_{VF}$, of all particles from extra 3VF, the threshold corrections, given in Eq.~(\ref{eq:Ti}), have the same value  in the leading log approximation, $T_i = (4/\pi) \ln (M_{VF}/M_Z)$, which is $\sim 6$ for $M_{VF} = 10$ TeV. These modify  gauge couplings at the EW scale by factor $1 + \alpha_i (M_Z) T_i$, which represents $\sim 10\%$, $\sim 20\%$, and $\sim 40\%$ corrections to gauge couplings $\alpha_1$, $\alpha_2$, and $\alpha_3$ respectively.  For our example with $\alpha_G = 0.3$ at $M_G = 2 \times 10^{16}$ GeV, the common mass $M_{VF} = 10$ TeV  leads to gauge couplings that are in agreement with experimental values with better than 8\% precision. The effect of $M_{VF} = 10$ TeV is shown in Fig.~\ref{fig:1} (bottom). This prediction is highly insensitive to the  value of $\alpha_G$. From Eqs.~(\ref{eq:sol_1loop}) and (\ref{eq:al3_2loop}) we can see that any $\alpha_G \gtrsim 0.3 $ contributes less than $\sim 10\%$ to gauge couplings at the EW scale. Its exact value is important only for precise prediction for gauge couplings as are exact masses of all particles from extra 3VF in case they are split.
Thus the only two relevant parameters are $M_G$ and $M_{VF}$, and with better than $8\%$ agreement of predicted values of three gauge couplings with experimental values, the predictive power of this scenario is comparable to that of minimal supersymmetric grand unification (which predicts 
 $\sim10\%$ too large $\alpha_3 (M_Z)$; this discrepancy is usually explained by 3\% - 4\% 
GUT scale threshold corrections, or by splitting  superpartner masses~\cite{PDG_GUTs}).

\noindent
{\bf Sensitivity to fundamental parameters.} 
Since $\alpha_G \gtrsim 0.3 $ contributes less than $\sim10\%$ to the EW scale values of gauge couplings,
changing it by a factor of 2 does not modify predicted values of gauge couplings by more than $\sim10\%$. This can be contrasted with high sensitivity in the minimal supersymmetric unification in which $\alpha_G^{-1} \simeq 24$, and $\sim3$ times larger value of $\alpha_3  (M_Z) $ compared to $\alpha_G$ is the result of a cancellation between the $\ln (M_G/M_Z)$ and 
$\alpha_G^{-1}$ terms in Eq.~(\ref{eq:sol_1loop}). Thus a given variation, $x\%$, of  $\alpha_G$ results in $\sim 3x\%$ variation of $\alpha_3  (M_Z) $.

Sensitivity of predicted gauge couplings to $M_G$ and $M_{VF}$ is also very small. Changing $M_G$ by a factor of two changes  $\ln (M_G/M_Z)$ term  in Eqs.~(\ref{eq:sol_1loop}) and (\ref{eq:al3_2loop}) and thus the weak scale values of gauge couplings by $\sim 2\%$. In minimal supersymmetric unification the same change would result in $\sim 6\%$ change in $\alpha_3  (M_Z) $ again as a result of the above mentioned cancellation.

Finally, changing $M_{VF}$ by a factor of two changes threshold corrections $T_i$ by $\sim 15\%$ which in turn changes EW scale values of gauge couplings by no more than $6\%$ (threshold corrections represent $\sim 40\%$ of $\alpha_3  (M_Z) $). 
 {\it Thus, changing any of the fundamental parameters by a factor of 2 does not modify predicted values of gauge couplings by more than $\sim$10\%.}

\noindent
{\bf Realistic example.} In order to obtain gauge couplings within experimental uncertainties, the masses of particles from extra 3VF must be split. This is indicated in Fig.~\ref{fig:1} (top) by close but not identical  scales at which the RG evolutions of gauge couplings in the SM and SM+3VF cross. These  crossing scales are $M_1\simeq 100$ TeV, $M_2 \simeq 1$ TeV, and $M_3 \simeq 10$ TeV for $\alpha_1$, $\alpha_2$, and $\alpha_3$ respectively. Thus any spectrum of particles from 3VF for which the threshold corrections given in Eq.~(\ref{eq:Ti}) equal to $T_i = (4/\pi) \ln (M_{i}/M_Z)$, as if all particles charged under given gauge symmetry had common mass $M_i$, will reproduce the measured values of gauge couplings. 

There are many solutions available. For simplicity, we present one that assumes universal masses for all vector-like pairs with the same quantum numbers. We use the same names for vector-like paris as for quarks and leptons in the SM, {\it e.g.} $m_Q$ is the common mass of vector-like quark doublets.
For $\alpha_G = 0.3$ at $M_G = 2 \times 10^{16}$ GeV a fully realistic example is obtained with $m_Q = 500$ GeV, $m_L = 95$ TeV, $m_U = 220$ TeV, $m_D = 180$ TeV, $m_E = 250$ TeV, and is shown in  Fig.~\ref{fig:2} (top). These parameters, specified to two significant figures, reproduce measured values of gauge couplings  with much better than 0.1\% precision and thus within small fractions of current experimental uncertainties. This is another way to demonstrate high insensitivity of gauge couplings to fundamental parameters.

\begin{figure}[t,h]
\includegraphics[width=2.6in]{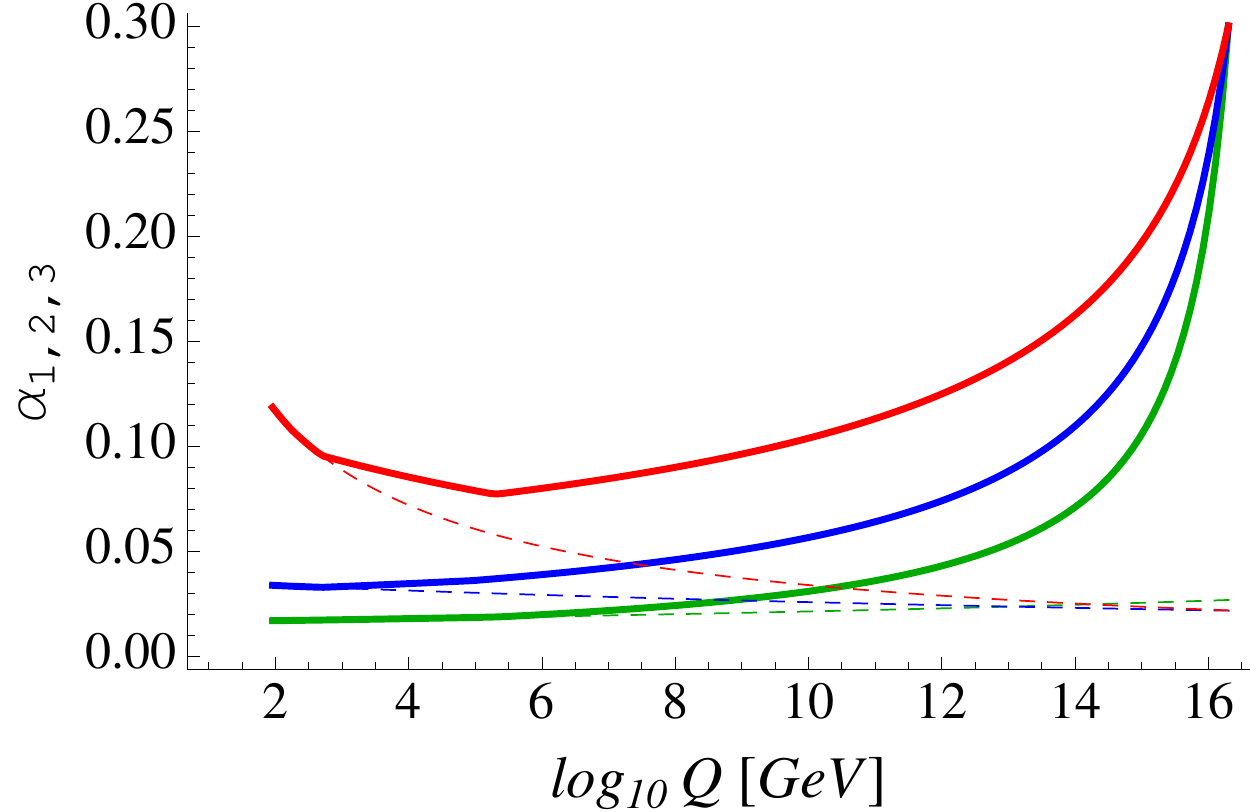}
\includegraphics[width=2.6in]{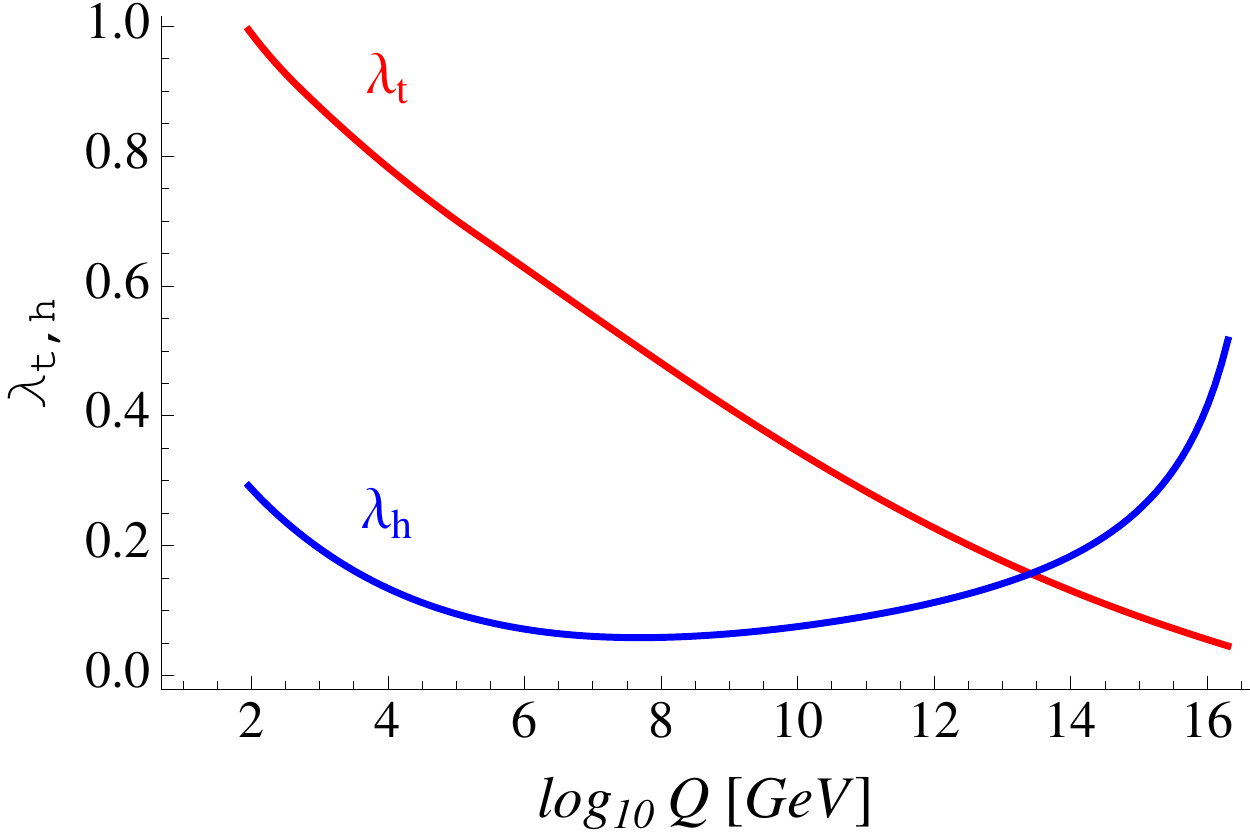}
\caption{Top: the same as in Fig.~\ref{fig:1} with masses of three pairs of vector-like fermions fixed to: $m_Q = 500$ GeV, $m_L = 95$ TeV,  $m_U = 220$ TeV, $m_D = 180$ TeV, and $m_E = 250$ TeV. Bottom: renormalization group running of the top Yukawa coupling, $\lambda_t$, and the Higgs quartic coupling, $\lambda_h$.}
\label{fig:2}
\end{figure}

For completeness, evolutions of the top Yukawa and  the Higgs quartic couplings are given in Fig.~\ref{fig:2} (bottom).  The Higgs quartic coupling remains positive all the way to the GUT scale and thus the electroweak minimum of the Higgs potential is stable.  This result holds for any Higgs boson mass in the currently allowed range~\cite{ATLAS:2012ae}.

\noindent
 {\bf Discussion and conclusions.}
 Since the effect of integrating out vector-like families can only increase gauge couplings at the EW scale,  see Eqs.~(\ref{eq:bi}-\ref{eq:Ti}) and Fig.~\ref{fig:1}, 
  values of predicted couplings without considering mass effect of 3VF have to be smaller than the measured values. In other words, the lines representing evolution of gauge couplings in the SM+3VF have to cross those of  the SM before they reach the $M_Z$ scale.
  This leads to 
a lower  limit on the GUT scale. % in order to achieve insensitive unification.
 For $\alpha_G = 0.3$ 
 %the lower limit on the  unification scale is 
 %the one-loop RGE's would require $M_G \gtrsim 6 \times 10^{16}$ GeV, two loop RGEs lower the minimal unification scale to  %$M_G \gtrsim 2 \times 10^{14}$ GeV. - this is just from alpha2 - no realistic unification.
the lower limit is  $M_G \gtrsim 1\times 10^{15}$ GeV. There is no upper limit on $M_G$, and it can be identified with the string scale or the planck scale. Increasing $M_G$ however increases the average mass scale of 3VF, that can be inferred from Eq.~(\ref{eq:sol_1loop}) or  Fig.~\ref{fig:1}, and also increases the sensitivity of predicted gauge couplings to $M_{VF}$. We checked numerically that increasing $M_G$ also requires larger splitting of masses of particles from 3VF.

%Insensitive unification can be also achieved with more than 3 vector-like families. Our example is the minimal one that makes $\alpha_3$ asymptotically divergent. 
Adding more vector-like families leads to faster running, and correspondingly the masses of VF must be larger or the GUT scale must be significantly lowered. Lowering the GUT scale is disfavored by proton decay limits, and increasing masses of VF (for example, with 4VF the average mass scale of VF increases to $\sim10^7$ GeV for $M_G = 2 \times 10^{16}$ GeV) leads to larger sensitivity 
of predicted gauge couplings to $M_{VF}$.  Thus, the SM+3VF represents the minimal and the most attractive scenario for insensitive unification. 

Although there are many possible  arrangements for masses of particles from 3VF that lead to precise predictions for gauge couplings, the existence of a solution with relatively small splitting between masses and their proximity to the EW scale is not  trivial. 
%Two or three orders of magnitude splitting between masses of particles from 3VF should not be surprising given the fact that there is more than 5 orders of magnitude hierarchy in masses of charged fermions in the SM. 
Requiring the smallest sensitivity to masses of particles from 3VF  suggests that
some of the extra fermions might be within the reach of the LHC. Since $\alpha_2$ has the lowest crossing scale, this applies especially to  quark doublets $Q_{L,R}$ and lepton doublets $L_{L,R}$. However, we checked numerically that breaking mass universality among particles with the same quantum numbers, each particle separately can be at current experimental limits.
% (the change in the threshold effect can be compensated by increased masses of other particles). 
A lot of freedom in masses of extra VF indicates that it should be  possible to  combine  explanations of various anomalies mentioned in the introduction with insensitive unification. It might also allow for simple GUT scale boundary conditions for these masses (this freedom is further enhanced by possible Yukawa couplings that change their RG evolution). Alternatively, the masses of 3VF might originate from Yukawa couplings to a singlet scalar field, or the extra 3VF might be charged under additional symmetries. 
%This will be explored elsewhere.

Similar unification mechanism is expected to work in  other models that do not modify  RGEs for gauge couplings dramatically, for example the two Higgs doublet model or other simple extensions of the SM. 

Another possible interpretation that this scenario offers is a non-perturbative unification at a high scale, as envisioned in Ref.~\cite{Maiani:1977cg}. However, it also allows an interpretation without any GUTs: gauge couplings might originate from random large values at a high scale. This follows from the fact that $ \alpha_i (M_Z)$ very weakly depend on  $\alpha_i(M_G)$, and thus whether gauge couplings really unify or not is not important for understanding the EW scale values of gauge couplings. However, this would  not provide an explanation of quantum numbers of SM particles which is perhaps the main virtue of simple GUTs.

The scenario we present  does not offer  an explanation of the hierarchy between the EW scale and the GUT scale. This might not be a disadvantage, since we have not discovered any  signs of new physics that would solve the hierarchy problem yet. 
%However, this simple extension of the SM provides an understanding of gauge couplings  that can be easily embedded into grand unified theories or new physics at the string or planck scale.

%%%%%%%%%%%%%%%%%%%%%%%%%%%%%%%%%%%%%%%%%%%%%%%%%%%%%%%%
%\acknowledgements
%%%%%%%%%%%%%%%%%%%%%%%%%%%%%%%%%%%%%%%%%%%%%%%%%%%%%%%%

\vspace{0.05cm}
\noindent
{\it Acknowledgments:} RD thanks H.D. Kim and  A. Wingerter for useful discussions, and SNU for kind hospitality during final stages of this project.

%%%%%%%%%%%%%%%%%%%%%%%%%%%%%%%%%%%%%%%%%%%%%%%%%%%%%%%%

%%%%%%%%%%%%%%%%%%%%%%%%%%%%%%%%%%%%%%%%%%%%%%%%%%%%%%%%%

%%%%%%%%%%%%%%%%%%%%%%%%%%%%%%%%%%%%%%%%%%%%%%%%%%%%%%%%%
\end{document}